# Scattering of electromagnetic wave from Young's pinholes upon a quasi-homogeneous anisotropic medium

**Josh Abbey[*] and Lin Frank Chan**

*Department of Physics, Kansas State University, Manhattan, KS 66506, USA*

[*]*Corresponding author: j.abbey@phys.ksu.edu*

**Abstract:** The scattering of an electromagnetic plane wave from Young's pinholes upon a quasi-homogeneous (QH) anisotropic medium is studied within the accuracy of the first-order Born approximation. By assuming that the diffractive wave from pinholes are paraxial with respect to the propagation direction, it is shown that the correlation of incident wave, anisotropic correlation statistics of the medium, and the pinhole size configuration can influence the statistical properties of the scattered field. In particular, the normalized spectral density, spectral degree of coherence, and spectral degree of polarization of the scattered field are derived by analytic forms, respectively.

## Introduction

The Quasi-homogeneous (QH) random medium, which was initially introduced by R. A. Silverman et al. two decades ago, revealing that the relation between the scattered field properties and the scattering potential profiles are closely related to each other [1]. It has been shown that the strength of the scattering potential of a QH medium varies much slower than the normalized correlation coefficient about their arguments at all frequency range of the field. The most representative discovery about this topic was announced by E. Wolf and his colleagues who proposed the scattering rule of light waves from a QH medium [2]. Since then, substantial research interest has been paid to the statistical characteristics of light scattering upon the QH medium. For instance, Fischer et al. showed that two-point spatial correlation function of a QH medium can be determined from the cross-spectral density (CSD) of the far-zone scattered field by specific experiments [3]. Further extension of the method was made for reconstructing the second-order statistical properties of light scattered from a QH random medium, assuming that the incident wave was the plane-wave pulse [4]. The application of QH random objects to the diffraction tomography revealed that the spatial degree of correlation of a QH medium can be determined by knowing the CSD of the scattered field [5]. It was further known that the second-order statistical properties of a radiated or scattered field, including the spectral density (SD) and spectral degree of coherence (SDOC), satisfy the reciprocity relations [6]. These relations were further extended to the electromagnetic cases which indicated that both the SD and SDOC of the scattered field from the electromagnetic plane wave on incidence must satisfy the reciprocity relations analogous to those in the scalar case [7]. The spectrum of electromagnetic wave scattered from a QH medium was studied in [8]. The correlation between intensity fluctuations of light scattered from a QH medium was studied for both the scalar and electromagnetic incident wave cases, respectively [9-11]. In addition to above studies, the scattering theory of light from a Gaussian-correlated, QH anisotropic medium was modeled by using the tensor method [12]. Also, analytical expressions were derived for revealing the reciprocity relations of a polychromatic plane wave scattered from a Gaussian-correlated, quasi-homogeneous, anisotropic medium [13]. The spectral shifts of light scattered from a QH anisotropic medium were quantitatively analyzed, while the results indicated that the spectral shift of the scattered field occurs when rotating the anisotropic medium [14]. The SDOC and degree of polarization (DOP) of an electromagnetic wave scattered from a QH anisotropic medium was analyzed in great detail, respectively [15]. It was shown that the DOP of an electromagnetic plane wave can convert from the uniform distribution to non-uniformly depolarized profile provided that the incident wave scatters from a QH anisotropic medium [16].

The above analyses mainly focused on the statistical properties of the scattered field from either a deterministic or random medium. In practical cases incident light beams on scattering always encounters obstacles along scattering paths. Statistical properties of truncated light waves scattering from specific medium must be obtained to achieve this effect. Li et al. have devoted major contributions in this field and introduced a novel scattering system incorporating Young's pinholes and a random medium. They derived reciprocity relations of the scattered field from the complex scattering system [17], and further extended the investigations to the fourth-order correlation statistics of the scattered field [18]. Furthermore, Li et al. also proposed an effective method that has shown applicable to determine the correlation function of a spatially QH medium [19]. In addition, they derived a set of sufficient conditions for letting the scattered spectrum be the same as that of the incident plane wave [20]. These major breakthroughs done by Li et al. regarding to this topic have greatly attracted our research interest, and enlightened us to utilize their models to further analyze the far-zone scattered properties of an electromagnetic plane wave scattering from a QH medium. Based upon the theoretical model of the cross-spectral density (CSD) from [17-20], we formulate the normalized spectral density, the DOC and DOP of scattered field which is produced by the scattering of a Young's diffractive electromagnetic wave from QH anisotropic medium. Based on the derived formulas, we can conclude that statistical moments of scattered field depend on both the vectorial correlations of incident waves and anisotropic spatial correlations of QH medium.

## Cross-spectral density matrix of scattered field generated by the scattering of Young's diffractive electromagnetic waves from QH anisotropic medium

An electromagnetic plane wave where the direction of propagation is represented by the unit vector $\hat{\mathbf{s}}_0$ is initially assumed to exist. It is also assumed the incident electromagnetic wave is truncated by Young's double pinholes, as presented in Fig. 1. After transmitting through the pinholes, the diffractive electromagnetic wave subsequently scatters on a QH anisotropic medium. Investigation will observe the second-order statistical properties, for instance the spectral density, DOC and DOP of scattered field. At the Young's opaque screen $A$, $Q(\hat{\boldsymbol{\rho}}_{1\perp})$ and $Q(\hat{\boldsymbol{\rho}}_{2\perp})$ denote the spatial locations of pinholes on an opaque screen $A$. $\hat{\boldsymbol{\rho}}_{1\perp}$ and $\hat{\boldsymbol{\rho}}_{2\perp}$ are the position vectors of the incident electromagnetic plane wave. $\phi_1, \phi_2$ represent the azimuthal angles between the position vectors $\hat{\boldsymbol{\rho}}_1, \hat{\boldsymbol{\rho}}_2$ and the unit vector $\hat{\mathbf{s}}_0$, respectively. The length $d$ is the spacing between pinholes. The electric fields of incident polychromatic electromagnetic plane wave are determined by

$$\mathbf{U}_\alpha^{(i)}(\hat{\boldsymbol{\rho}}, \omega) = a_\alpha(\omega) \exp(-ik\hat{\mathbf{s}}_0 \cdot \hat{\boldsymbol{\rho}}), \quad (\alpha=x, y), \tag{1}$$

where the superscript (*i*) represents the electric field components of incident electromagnetic waves along the $\alpha$ direction. $a_\alpha(\omega)$ is, in general, complex, and $k=\omega/c$, with $c$ being the speed of light in vacuum. We also assume the transmitted light wave subsequently propagates and further interacts with a QH anisotropic medium at the two points $P(\hat{\mathbf{r}}_1')$ and $P(\hat{\mathbf{r}}_2')$. On the basis of the elementary theory of electromagnetic waves transmitting through Young's pinholes, the resultant electric field components of the diffractive light waves specified at the points $P(\hat{\mathbf{r}}_1')$ and $P(\hat{\mathbf{r}}_2')$ can be represented as the analytical forms [21-24]

$$\mathbf{U}_\alpha^{(t)}(\hat{\mathbf{r}}_1',\omega) = -\frac{i}{\lambda}dS\left[\frac{\exp(ikR_{11})}{R_{11}}\mathbf{U}_\alpha^{(i)}(\hat{\boldsymbol{\rho}}_1,\omega) + \frac{\exp(ikR_{12})}{R_{12}}\mathbf{U}_\alpha^{(i)}(\hat{\boldsymbol{\rho}}_2,\omega)\right], (\alpha=x, y), \qquad (2)$$

$$\mathbf{U}_\alpha^{(t)}(\hat{\mathbf{r}}_2',\omega) = -\frac{i}{\lambda}dS\left[\frac{\exp(ikR_{21})}{R_{21}}\mathbf{U}_\alpha^{(i)}(\hat{\boldsymbol{\rho}}_1,\omega) + \frac{\exp(ikR_{22})}{R_{22}}\mathbf{U}_\alpha^{(i)}(\hat{\boldsymbol{\rho}}_2,\omega)\right], (\alpha=x, y), \qquad (3)$$

Where, $dS$ represents the area of each pinhole, $\mathbf{U}_\alpha^{(i)}(\hat{\boldsymbol{\rho}}_j,\omega)$ ($j$=1, 2) has been determined by Eq. (1). The superscript ($t$) in Eq. (2) and (3) denotes the electric field components of transmitted light waves. $R_{mn}$ ($m$, $n$=1 or 2) are the geometrical lengths between the points $Q(\hat{\boldsymbol{\rho}}_{m\perp})$ and $P(\hat{\mathbf{r}}_n')$, as presented in Fig. 1. Recalling the expression for the elements of the cross-spectral density matrix of an electromagnetic wave [25]

$$\mathbf{W}_{\alpha\beta}^{(i)}(\hat{\boldsymbol{\rho}}_1,\hat{\boldsymbol{\rho}}_2,\omega) = \left\langle \mathbf{U}_\alpha^{(i)*}(\hat{\boldsymbol{\rho}}_1,\omega)\mathbf{U}_\beta^{(i)}(\hat{\boldsymbol{\rho}}_2,\omega) \right\rangle, \quad (\alpha=x, y; \beta=x, y), \qquad (4)$$

where the angular bracket represents the ensemble average taken over the electric field and the asterisk denotes the complex conjugation. In this paper, we assume the two pinholes at the opaque screen $A$ can be considered as symmetric with respect to the z axis. Accordingly, one may readily obtain the following relation

$$\left|\hat{\boldsymbol{\rho}}_j\right|\cos\phi_j = R', \quad (j=1, 2); \qquad (5)$$

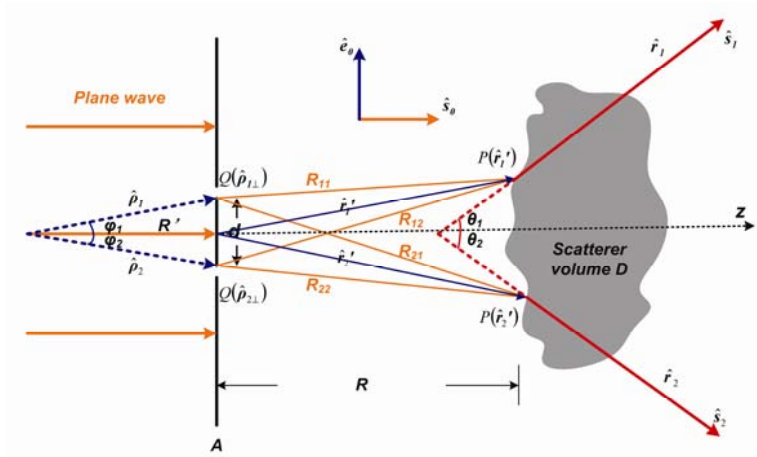

Fig. 1 Schematic diagram for a Young's diffractive electromagnetic plane wave scattering on a QH anisotropic medium. $Q(\hat{\boldsymbol{\rho}}_1)$ and $Q(\hat{\boldsymbol{\rho}}_2)$ constitute the Young's pinholes locating at the opaque screen $A$. $P(\hat{\mathbf{r}}_1')$ and $P(\hat{\mathbf{r}}_2')$ are the two points where the transmitted light interacts with QH anisotropic medium. $d$ denotes the spacing between pinholes, $R$ represents the distance between the opaque screen and scatterer. $\theta_1$ and $\theta_2$ are the scattering angles corresponding to the unit vectors $\hat{\mathbf{s}}_1$ and $\hat{\mathbf{s}}_2$, respectively. $\hat{\mathbf{e}}_0$ can be introduced as the unit vector which remains perpendicular to the wave vector $\hat{\mathbf{s}}_0$.

Assuming that the length between the opaque screen $A$ and the scatterer is large enough and the transmitted light wave from the pinholes can be approximated as the paraxial beam; the paraxial approximation may be correspondingly calculated [21]

$$R_{mn} \approx R, \quad (m=1, 2; n=1, 2), \tag{6}$$

$$R_{21} - R_{11} \approx -\frac{(\hat{\mathbf{r}}_1' + \hat{\mathbf{r}}_2') \cdot \hat{\mathbf{e}}_0}{2R} d, \tag{7}$$

$$R_{22} - R_{12} \approx -\frac{(\hat{\mathbf{r}}_1' + \hat{\mathbf{r}}_2') \cdot \hat{\mathbf{e}}_0}{2R} d, \tag{8}$$

$$R_{12} - R_{11} \approx \frac{\hat{\mathbf{r}}_1' \cdot \hat{\mathbf{e}}_0}{R} d, \tag{9}$$

$$R_{22} - R_{21} \approx -\frac{\hat{\mathbf{r}}_2' \cdot \hat{\mathbf{e}}_0}{R} d, \tag{10}$$

where $\hat{\mathbf{e}}_0$ can be introduced as the unit vector of which the direction is perpendicular that of $\hat{\mathbf{s}}_0$, as presented in Fig. 1. Accordingly, the following relation can be established between the above two unit vectors

$$\hat{\mathbf{e}}_0 \cdot \hat{\mathbf{s}}_0 = 0, \tag{11}$$

By substituting from Eqs. (2)-(3) into (4) and utilizing (5)-(11), Li et al. have introduced the CSD matrix of the transmitted wave from pinholes, giving by the following form [17-20]

$$\mathbf{W}_{\alpha\beta}^{(t)}(\hat{\mathbf{r}}_1', \hat{\mathbf{r}}_2', \omega) = \left(\frac{dS}{\lambda R}\right)^2 \langle a_\alpha^*(\omega) a_\beta(\omega) \rangle \left\{ 2\exp\left[-i\frac{kd}{2R}(\hat{\mathbf{r}}_1' + \hat{\mathbf{r}}_2') \cdot \hat{\mathbf{e}}_0\right] + \exp\left[-i\frac{kd}{2R}(\hat{\mathbf{r}}_2' - \hat{\mathbf{r}}_1') \cdot \hat{\mathbf{e}}_0\right] + \exp\left[i\frac{kd}{2R}(\hat{\mathbf{r}}_2' - \hat{\mathbf{r}}_1') \cdot \hat{\mathbf{e}}_0\right] \right\},$$

$$(\alpha=x, y; \beta=x, y). \tag{12}$$

In this paper, we utilize Eq. (12) to perform further analysis on scattering of transmitted wave upon a spatially QH anisotropic medium with a finite volume $D$. The scattering of the transmitted light upon the scatterer may be considered sufficiently weak enough to allow the first-order Born approximation to be utilized for treating the entire scattering progress. To achieve our goal of calculating the scattered field components in the far field, we recall the scattering integral forms introduced in [18, 26, 27]

$$\mathbf{U}_x^{(s)}(r\hat{\mathbf{s}},\omega) = \frac{\exp(ikr)}{r} \int_D \mathbf{U}_x^{(t)}(\hat{\mathbf{r}}',\omega) \mathbf{F}_x(\hat{\mathbf{r}}',\omega) \exp(-ik\hat{\mathbf{s}}\cdot\hat{\mathbf{r}}') d^3\hat{\mathbf{r}}', \tag{13}$$

$$\mathbf{U}_y^{(s)}(r\hat{\mathbf{s}},\omega) = \frac{\exp(ikr)}{r} \cos^2\theta \int_D \mathbf{U}_y^{(t)}(\hat{\mathbf{r}}',\omega) \mathbf{F}_y(\hat{\mathbf{r}}',\omega) \exp(-ik\hat{\mathbf{s}}\cdot\hat{\mathbf{r}}') d^3\hat{\mathbf{r}}', \tag{14}$$

$$\mathbf{U}_z^{(s)}(r\hat{\mathbf{s}},\omega) = -\frac{\exp(ikr)}{r} \sin\theta \cos\theta \int_D \mathbf{U}_y^{(t)}(\hat{\mathbf{r}}',\omega) \mathbf{F}_z(\hat{\mathbf{r}}',\omega) \exp(-ik\hat{\mathbf{s}}\cdot\hat{\mathbf{r}}') d^3\hat{\mathbf{r}}', \tag{15}$$

Where the unit vector $\hat{\mathbf{s}}$ denotes the direction along which the scattered light propagates, $\mathbf{F}_j(\hat{\mathbf{r}}',\omega)$ ($j=x, y, z$) is the diagonal element of scattering potential of QH anisotropic medium, $\theta$ represents the scattering angle, the subscript $D$ appeared in Eqs. (13)-(15) amounts to the performance by taking the integrations over the spatial volume of QH anisotropic medium. It is essential to note that Eqs. (13)-(15) are reached by using the model of the anisotropic medium from [18, 26, 27] where a 3×3 scattering matrix with only diagonal elements was employed to define the scattering potential effect. Such definition of the scattering matrix provides an effective approach for us to further obtain second-order statistical properties of the far-zone scattered field, e.g. scattered intensity, SDOC and SDOP. By substituting from Eqs. (13)-(15) into (4), one may obtain the elements of the cross-spectral density matrix of scattered field along arbitrary scattering paths $r\hat{\mathbf{s}}_1$ and $r\hat{\mathbf{s}}_2$

$$\mathbf{W}_{xx}^{(s)}(r\hat{\mathbf{s}}_1, r\hat{\mathbf{s}}_2,\omega) = \frac{|\gamma(\omega)|^2}{|\gamma(\omega)|^2+1} S^{(i)}(\omega) \left(\frac{dS}{\lambda Rr}\right)^2 \mathbf{\Gamma}_{xx}(\hat{\mathbf{s}}_1,\hat{\mathbf{s}}_2,\omega), \tag{16}$$

$$\mathbf{W}_{xy}^{(s)}(r\hat{\mathbf{s}}_1, r\hat{\mathbf{s}}_2,\omega) = \frac{|\gamma(\omega)|}{|\gamma(\omega)|^2+1} S^{(i)}(\omega) \left(\frac{dS}{\lambda Rr}\right)^2 \cos^2\theta_2 \mathbf{\Gamma}_{xy}(\hat{\mathbf{s}}_1,\hat{\mathbf{s}}_2,\omega), \tag{17}$$

$$\mathbf{W}_{xz}^{(s)}(r\hat{\mathbf{s}}_1, r\hat{\mathbf{s}}_2,\omega) = -\frac{|\gamma(\omega)|}{|\gamma(\omega)|^2+1} S^{(i)}(\omega) \left(\frac{dS}{\lambda Rr}\right)^2 \sin\theta_2 \cos\theta_2 \mathbf{\Gamma}_{xz}(\hat{\mathbf{s}}_1,\hat{\mathbf{s}}_2,\omega), \tag{18}$$

$$\mathbf{W}_{yy}^{(s)}(r\hat{\mathbf{s}}_1, r\hat{\mathbf{s}}_2,\omega) = \frac{1}{|\gamma(\omega)|^2+1} S^{(i)}(\omega) \left(\frac{dS}{\lambda Rr}\right)^2 \cos^2\theta_1 \cos^2\theta_2 \mathbf{\Gamma}_{yy}(\hat{\mathbf{s}}_1,\hat{\mathbf{s}}_2,\omega), \tag{19}$$

$$\mathbf{W}_{yz}^{(s)}(r\hat{\mathbf{s}}_1, r\hat{\mathbf{s}}_2, \omega) = -\frac{1}{|\gamma(\omega)|^2 + 1} S^{(i)}(\omega) \left(\frac{dS}{\lambda Rr}\right)^2 \cos^2\theta_1 \sin\theta_2 \cos\theta_2 \mathbf{\Gamma}_{yz}(\hat{\mathbf{s}}_1, \hat{\mathbf{s}}_2, \omega), \tag{20}$$

$$\mathbf{W}_{zz}^{(s)}(r\hat{\mathbf{s}}_1, r\hat{\mathbf{s}}_2, \omega) = \frac{1}{|\gamma(\omega)|^2 + 1} S^{(i)}(\omega) \left(\frac{dS}{\lambda Rr}\right)^2 \sin\theta_1 \cos\theta_1 \sin\theta_2 \cos\theta_2 \mathbf{\Gamma}_{zz}(\hat{\mathbf{s}}_1, \hat{\mathbf{s}}_2, \omega), \tag{21}$$

with,

$$S^{(i)}(\omega) = \langle |a_x(\omega)|^2 \rangle + \langle |a_y(\omega)|^2 \rangle, \tag{22}$$

as the spectral density of the incident electromagnetic plane waves. The proportional factor $\gamma(\omega)$ in Eqs. (16)-(21) depends only on frequency and may be calculated by

$$a_x(\omega) = \gamma(\omega) a_y(\omega), \tag{23}$$

and $\mathbf{\Gamma}_{\alpha\beta}(\hat{\mathbf{s}}_1, \hat{\mathbf{s}}_2, \omega)$ may be further expressed as the integral form

$$\mathbf{\Gamma}_{\alpha\beta}(\hat{\mathbf{s}}_1, \hat{\mathbf{s}}_2, \omega) = \int_D \int_D \langle \mathbf{F}_\alpha^*(\hat{\mathbf{r}}_1', \omega) \mathbf{F}_\beta(\hat{\mathbf{r}}_2', \omega) \rangle \left\{ 2\exp\left[-i\frac{k}{2}\left(\hat{\mathbf{s}}_2 - \hat{\mathbf{s}}_1 + \frac{d}{R}\hat{\mathbf{e}}_0\right)\cdot(\hat{\mathbf{r}}_1' + \hat{\mathbf{r}}_2')\right.\right.$$

$$\left.-i\frac{k}{2}(\hat{\mathbf{s}}_1 + \hat{\mathbf{s}}_2)\cdot(\hat{\mathbf{r}}_2' - \hat{\mathbf{r}}_1')\right] + \exp\left[-i\frac{k}{2}(\hat{\mathbf{s}}_2 - \hat{\mathbf{s}}_1)\cdot(\hat{\mathbf{r}}_1' + \hat{\mathbf{r}}_2') - i\frac{k}{2}\left(\hat{\mathbf{s}}_1 + \hat{\mathbf{s}}_2 - \frac{d}{R}\hat{\mathbf{e}}_0\right)\cdot(\hat{\mathbf{r}}_2' - \hat{\mathbf{r}}_1')\right]$$

$$\left.+ \exp\left[-i\frac{k}{2}(\hat{\mathbf{s}}_2 - \hat{\mathbf{s}}_1)\cdot(\hat{\mathbf{r}}_1' + \hat{\mathbf{r}}_2') - i\frac{k}{2}\left(\hat{\mathbf{s}}_1 + \hat{\mathbf{s}}_2 + \frac{d}{R}\hat{\mathbf{e}}_0\right)\cdot(\hat{\mathbf{r}}_2' - \hat{\mathbf{r}}_1')\right]\right\} d^3\hat{\mathbf{r}}_1' d^3\hat{\mathbf{r}}_2', \tag{24}$$

Based on the definition of the scattering potential of a spatially anisotropic medium, the correlation function of a QH, anisotropic medium was introduced in [26, 27]

$$\langle \mathbf{F}_\alpha^*(\hat{\mathbf{r}}_1', \omega) \mathbf{F}_\beta(\hat{\mathbf{r}}_2', \omega) \rangle = \mathbf{S}_{\alpha\beta}^{(F)}[(\hat{\mathbf{r}}_1' + \hat{\mathbf{r}}_2')/2, \omega] \mathbf{\eta}_{\alpha\beta}^{(F)}(\hat{\mathbf{r}}_2' - \mathbf{r}_1', \omega), \quad (\alpha, \beta = x, y, z), \tag{25}$$

where $\mathbf{S}_{\alpha\beta}^{(F)}$ and $\mathbf{\eta}_{\alpha\beta}^{(F)}$ are the so-called strengths and normalized correlation coefficients of the scattering potential of QH anisotropic medium, respectively. By using Eqs. (24) and (25), we change the integral variables

$$\hat{\mathbf{R}}_s^+ = \hat{\mathbf{r}}_1' + \hat{\mathbf{r}}_2', \qquad \hat{\mathbf{R}}_s^- = \hat{\mathbf{r}}_2' - \hat{\mathbf{r}}_1', \tag{26}$$

and substitute from Eqs. (25) and (26) in (24), $\upsilon_{\alpha\beta}$ can be further rewritten as the following form

$$\mathbf{\Gamma}_{\alpha\beta}(\hat{\mathbf{s}}_1, \hat{\mathbf{s}}_2, \omega) = 2\tilde{\mathbf{S}}_{\alpha\beta}^{(F)}\left[k\left(\hat{\mathbf{s}}_2 - \hat{\mathbf{s}}_1 + \frac{d}{R}\hat{\mathbf{e}}_0\right)\right] \tilde{\mathbf{\eta}}_{\alpha\beta}^{(F)}\left[k(\hat{\mathbf{s}}_1 + \hat{\mathbf{s}}_2)/2\right]$$

$$+\tilde{\mathbf{S}}_{\alpha\beta}^{(F)}[k(\hat{\mathbf{s}}_2 - \hat{\mathbf{s}}_1)] \tilde{\mathbf{\eta}}_{\alpha\beta}^{(F)}\left[\frac{k}{2}\left(\hat{\mathbf{s}}_1 + \hat{\mathbf{s}}_2 + \frac{d}{R}\hat{\mathbf{e}}_0\right)\right] + \tilde{\mathbf{S}}_{\alpha\beta}^{(F)}[k(\hat{\mathbf{s}}_2 - \hat{\mathbf{s}}_1)] \tilde{\mathbf{\eta}}_{\alpha\beta}^{(F)}\left[\frac{k}{2}\left(\hat{\mathbf{s}}_1 + \hat{\mathbf{s}}_2 - \frac{d}{R}\hat{\mathbf{e}}_0\right)\right], \tag{27}$$

with,

$$\tilde{\mathbf{S}}_{\alpha\beta}^{(F)}(k\mathbf{u}) = \int_D \mathbf{S}_{\alpha\beta}^{(F)}(\hat{\mathbf{R}}_s^+)\exp(ik\hat{\mathbf{R}}_s^+ \cdot \mathbf{u})d^3\hat{\mathbf{R}}_s^+, \tag{28}$$

$$\tilde{\boldsymbol{\eta}}_{\alpha\beta}^{(F)}(k\mathbf{u}) = \int_D \boldsymbol{\eta}_{\alpha\beta}^{(F)}(\hat{\mathbf{R}}_s^-)\exp(ik\hat{\mathbf{R}}_s^- \cdot \mathbf{u})d^3\hat{\mathbf{R}}_s^-, \tag{29}$$

as the three-dimensional spatial Fourier transformation of $\mathbf{S}_{\alpha\beta}^{(F)}$ and $\boldsymbol{\eta}_{\alpha\beta}^{(F)}$, respectively. Except for Eqs. (16)-(21), the residual off-diagonal elements of the cross-spectral density matrix of the scattering potential, namely $\mathbf{W}_{yx}^{(s)}, \mathbf{W}_{zx}^{(s)}, \mathbf{W}_{zy}^{(s)}$ may be obtained as

$$\mathbf{W}_{yx}^{(s)}(r\hat{\mathbf{s}}_1, r\hat{\mathbf{s}}_2, \omega) = \mathbf{W}_{xy}^{(s)*}(r\hat{\mathbf{s}}_2, r\hat{\mathbf{s}}_1, \omega) = \frac{|\gamma(\omega)|}{|\gamma(\omega)|^2 + 1} S^{(i)}(\omega)\left(\frac{dS}{\lambda Rr}\right)^2 \cos^2\theta_1 \boldsymbol{\Gamma}_{xy}^*(\hat{\mathbf{s}}_2, \hat{\mathbf{s}}_1, \omega), \tag{30}$$

$$\mathbf{W}_{zx}^{(s)}(r\hat{\mathbf{s}}_1, r\hat{\mathbf{s}}_2, \omega) = \mathbf{W}_{xz}^{(s)*}(r\hat{\mathbf{s}}_2, r\hat{\mathbf{s}}_1, \omega) = -\frac{|\gamma(\omega)|}{|\gamma(\omega)|^2 + 1} S^{(i)}(\omega)\left(\frac{dS}{\lambda Rr}\right)^2 \sin\theta_1\cos\theta_1 \boldsymbol{\Gamma}_{xz}^*(\hat{\mathbf{s}}_2, \hat{\mathbf{s}}_1, \omega), \tag{31}$$

$$\mathbf{W}_{zy}^{(s)}(r\hat{\mathbf{s}}_1, r\hat{\mathbf{s}}_2, \omega) = \mathbf{W}_{yz}^{(s)*}(r\hat{\mathbf{s}}_2, r\hat{\mathbf{s}}_1, \omega) = -\frac{1}{|\gamma(\omega)|^2 + 1} S^{(i)}(\omega)\left(\frac{dS}{\lambda Rr}\right)^2 \sin\theta_1\cos\theta_1\cos^2\theta_2 \boldsymbol{\Gamma}_{yz}^*(\hat{\mathbf{s}}_2, \hat{\mathbf{s}}_1, \omega), \tag{32}$$

with the complex conjugation $\boldsymbol{\Gamma}_{\alpha\beta}^*$ provided by

$$\boldsymbol{\Gamma}_{\alpha\beta}^*(\hat{\mathbf{s}}_2, \hat{\mathbf{s}}_1, \omega) = 2\tilde{\mathbf{S}}_{\alpha\beta}^{(F)}\left[k\left(\hat{\mathbf{s}}_2 - \hat{\mathbf{s}}_1 - \frac{d}{R}\hat{\mathbf{e}}_0\right)\right]\tilde{\boldsymbol{\eta}}_{\beta\alpha}^{(F)}\left[k(\hat{\mathbf{s}}_1 + \hat{\mathbf{s}}_2)/2\right] + \tilde{\mathbf{S}}_{\alpha\beta}^{(F)}\left[k(\hat{\mathbf{s}}_2 - \hat{\mathbf{s}}_1)\right]$$

$$\times \tilde{\boldsymbol{\eta}}_{\beta\alpha}^{(F)}\left[\frac{k}{2}\left(\hat{\mathbf{s}}_1 + \hat{\mathbf{s}}_2 + \frac{d}{R}\hat{\mathbf{e}}_0\right)\right] + \tilde{\mathbf{S}}_{\alpha\beta}^{(F)}\left[k(\hat{\mathbf{s}}_2 - \hat{\mathbf{s}}_1)\right]\tilde{\boldsymbol{\eta}}_{\beta\alpha}^{(F)}\left[\frac{k}{2}\left(\hat{\mathbf{s}}_1 + \hat{\mathbf{s}}_2 - \frac{d}{R}\hat{\mathbf{e}}_0\right)\right], \tag{33}$$

the deductions of Eqs. (30)-(33) are created based on the property that elements of the scattering potential of QH anisotropic medium should fulfill the following conditions

$$\tilde{\mathbf{S}}_{\alpha\beta}^{(F)}[k(\hat{\mathbf{s}}_2 - \hat{\mathbf{s}}_1)] = \tilde{\mathbf{S}}_{\beta\alpha}^{(F)}[k(\hat{\mathbf{s}}_2 - \hat{\mathbf{s}}_1)] = \left\{\tilde{\mathbf{S}}_{\alpha\beta}^{(F)}[k(\hat{\mathbf{s}}_1 - \hat{\mathbf{s}}_2)]\right\}^*, \tag{34}$$

$$\left\{\tilde{\boldsymbol{\eta}}_{\alpha\beta}^{(F)}\left[k\left(\frac{\hat{\mathbf{s}}_1 + \hat{\mathbf{s}}_2}{2} - \hat{\mathbf{s}}_0\right)\right]\right\}^* = \left\{\tilde{\boldsymbol{\eta}}_{\alpha\beta}^{(F)}\left[k\left(\hat{\mathbf{s}}_0 - \frac{\hat{\mathbf{s}}_1 + \hat{\mathbf{s}}_2}{2}\right)\right]\right\}^* = \tilde{\boldsymbol{\eta}}_{\beta\alpha}^{(F)}\left[k\left(\frac{\hat{\mathbf{s}}_1 + \hat{\mathbf{s}}_2}{2} - \hat{\mathbf{s}}_0\right)\right], \tag{35}$$

Eqs. (16)-(21) together with (30)-(32) constitute the 3×3 cross-spectral density matrix of scattered field provided the QH anisotropic medium is illuminated by electromagnetic waves which first transmit through Young's pinholes.

## Second-order moments of scattered field generated by the scattering of Young's diffractive electromagnetic waves from QH anisotropic medium

The second-order moments of scattered field, namely that the normalized spectral density, DOC and DOP of scattered light waves, are investigated. On the basis of Eqs. (16)-(21) and (30)-(32), the spectral density, DOC and DOP of scattered field may be separately deduced according to the *Unified Theory of Coherence and Polarization of Electromagnetic Waves* [28-30]

$$S^{(s)}(r\hat{\mathbf{s}},\omega) = Tr\left[W_{ij}^{(s)}(r\hat{\mathbf{s}},r\hat{\mathbf{s}},\omega)\right], \tag{36}$$

$$\mu^{(s)}(r\hat{\mathbf{s}}_1,r\hat{\mathbf{s}}_2,\omega) = \frac{Tr\left[\mathbf{W}_{\alpha\beta}^{(s)}(r\hat{\mathbf{s}}_1,r\hat{\mathbf{s}}_2,\omega)\right]}{\sqrt{Tr\left[\mathbf{W}_{\alpha\beta}^{(s)}(r\hat{\mathbf{s}}_1,r\hat{\mathbf{s}}_1,\omega)\right]}\sqrt{Tr\left[\mathbf{W}_{\alpha\beta}^{(s)}(r\hat{\mathbf{s}}_2,r\hat{\mathbf{s}}_2,\omega)\right]}}, \tag{37}$$

$$P^{(s)}(r\hat{\mathbf{s}},\omega) = \sqrt{\frac{3}{2}\left\{\frac{Tr\left[\mathbf{W}_{\alpha\beta}^{(s)}(r\hat{\mathbf{s}},r\hat{\mathbf{s}},\omega)^2\right]}{\left[Tr\mathbf{W}_{\alpha\beta}^{(s)}(r\hat{\mathbf{s}},r\hat{\mathbf{s}},\omega)\right]^2} - \frac{1}{3}\right\}}, \tag{38}$$

Where, *Tr* denotes performing the trace of the cross-spectral density matrix of scattered field. On substituting Eqs. (16), (19) and (21) into (36), the resultant spectral density of scattered field yields the following expression

$$S^{(s)}(r\hat{\mathbf{s}},\omega) = \frac{|\gamma(\omega)|^2}{|\gamma(\omega)|^2+1}S^{(i)}(\omega)\left(\frac{dS}{\lambda Rr}\right)^2\mathbf{\Gamma}_{\mathbf{xx}}(\hat{\mathbf{s}},\omega) + \frac{1}{|\gamma(\omega)|^2+1}S^{(i)}(\omega)\left(\frac{dS}{\lambda Rr}\right)^2\cos^4\theta\mathbf{\Gamma}_{\mathbf{yy}}(\hat{\mathbf{s}},\omega)$$

$$+\frac{1}{|\gamma(\omega)|^2+1}S^{(i)}(\omega)\left(\frac{dS}{\lambda Rr}\right)^2\sin^2\theta\cos^2\theta\mathbf{\Gamma}_{\mathbf{zz}}(\hat{\mathbf{s}},\omega), \tag{39}$$

with $\mathbf{\Gamma}_{\alpha\alpha}(\hat{\mathbf{s}},\omega)$ calculated by

$$\mathbf{\Gamma}_{\alpha\alpha}(\hat{\mathbf{s}},\omega) = 2\tilde{\mathbf{S}}_{\alpha\alpha}^{(F)}\left(\frac{kd}{R}\hat{\mathbf{e}}_0\right)\tilde{\boldsymbol{\eta}}_{\alpha\alpha}^{(F)}(k\hat{\mathbf{s}}) + \tilde{\mathbf{S}}_{\alpha\alpha}^{(F)}(0)\tilde{\boldsymbol{\eta}}_{\alpha\alpha}^{(F)}\left[k\left(\hat{\mathbf{s}}+\frac{d}{2R}\hat{\mathbf{e}}_0\right)\right] + \tilde{\mathbf{S}}_{\alpha\alpha}^{(F)}(0)\tilde{\boldsymbol{\eta}}_{\alpha\alpha}^{(F)}\left[k\left(\hat{\mathbf{s}}-\frac{d}{2R}\hat{\mathbf{e}}_0\right)\right], \tag{40}$$

utilizing the expression for the normalized spectral density of scattered field as the following expression [31, 32]

$$S_N^{(s)}(r\hat{\mathbf{s}},\omega) = \frac{S^{(s)}(r\hat{\mathbf{s}},\omega)}{\int_0^\infty S^{(s)}(r\hat{\mathbf{s}},\omega)d\omega}, \tag{41}$$

assuming further that proportional factor $\gamma(\omega)$ is independent of frequency i.e. $\gamma(\omega) \cong \gamma$, the resultant normalized spectral density of scattered field may be obtained by the substitution of Eq. (39) into (41), such that

$$S_N^{(s)}(r\hat{\mathbf{s}},\omega) = S_N^{(i)}(\omega)\bar{M}_u(\hat{\mathbf{s}},\omega), \tag{42}$$

with,

$$S_N^{(i)}(r\hat{\mathbf{s}},\omega) = \frac{S^{(i)}(\omega)}{\int_0^\infty S^{(i)}(\omega)d\omega}, \tag{43}$$

as the purportedly normalized spectral density of incident electromagnetic plane waves. In Eq. (42), $\bar{M}_u$ is the spectral modifier which may be expressed as

$$\bar{M}_u(\hat{\mathbf{s}},\omega) = \omega^2 \frac{\gamma \mathbf{\Gamma}_{xx}(\hat{\mathbf{s}},\omega) + \cos^4\theta \mathbf{\Gamma}_{yy}(\hat{\mathbf{s}},\omega) + \sin^2\theta \cos^2\theta \mathbf{\Gamma}_{zz}(\hat{\mathbf{s}},\omega)}{\gamma \int_0^\infty \omega^2 \mathbf{\Gamma}_{xx}(\hat{\mathbf{s}},\omega)d\omega + \cos^4\theta \int_0^\infty \omega^2 \mathbf{\Gamma}_{yy}(\hat{\mathbf{s}},\omega) + \sin^2\theta \cos^2\theta \int_0^\infty \omega^2 \mathbf{\Gamma}_{zz}(\hat{\mathbf{s}},\omega)}, \tag{44}$$

the DOC and DOP of scattered field similarly may be derived by substituting (39) into (37) and (38), respectively

$$\mu^{(s)}(r\hat{\mathbf{s}}_1, r\hat{\mathbf{s}}_2, \omega) = \frac{\gamma^2 \mathbf{\Gamma}_{xx}(\hat{\mathbf{s}}_1, \hat{\mathbf{s}}_2, \omega) + \cos^2\theta_1 \cos^2\theta_2 \mathbf{\Gamma}_{yy}(\hat{\mathbf{s}}_1, \hat{\mathbf{s}}_2, \omega) + \sin\theta_1 \cos\theta_1 \sin\theta_2 \cos\theta_2 \mathbf{\Gamma}_{zz}(\hat{\mathbf{s}}_1, \hat{\mathbf{s}}_2, \omega)}{\sqrt{\gamma^2 \mathbf{\Gamma}_{xx}(\hat{\mathbf{s}}_1, \omega) + \cos^4\theta_1 \mathbf{\Gamma}_{yy}(\hat{\mathbf{s}}_1, \omega) + \sin^2\theta_1 \cos^2\theta_1 \mathbf{\Gamma}_{zz}(\hat{\mathbf{s}}_1, \omega)}}$$
$$\times \frac{1}{\sqrt{\gamma^2 \mathbf{\Gamma}_{xx}(\hat{\mathbf{s}}_2, \omega) + \cos^4\theta_2 \mathbf{\Gamma}_{yy}(\hat{\mathbf{s}}_2, \omega) + \sin^2\theta_2 \cos^2\theta_2 \mathbf{\Gamma}_{zz}(\hat{\mathbf{s}}_2, \omega)}}, \tag{45}$$

$$P^{(s)}(r\hat{\mathbf{s}},\omega) = \sqrt{\frac{3}{2}\frac{\gamma^4 |\mathbf{\Gamma}_{xx}(\hat{\mathbf{s}},\omega)|^2 + \cos^8\theta |\mathbf{\Gamma}_{yy}(\hat{\mathbf{s}},\omega)|^2 + \sin^4\theta \cos^4\theta |\mathbf{\Gamma}_{zz}(\hat{\mathbf{s}},\omega)|^2}{[\gamma^2 \mathbf{\Gamma}_{xx}(\hat{\mathbf{s}},\omega) + \cos^4\theta \mathbf{\Gamma}_{yy}(\hat{\mathbf{s}},\omega) + \sin^2\theta \cos^2\theta \mathbf{\Gamma}_{zz}(\hat{\mathbf{s}},\omega)]^2} - \frac{1}{3}}, \tag{46}$$

## Discussion

In this section, we specifically compare our results with some previous ones from [17], which has provided a theoretical basis for deriving our current results. Therefore, it is rather necessary to compare our results with those from [17]. First, if we compare Eq. (39) with Eqs. (24) and (26) of [17], it is shown that the spectral density of scattered field derived in this paper is similar to but still different those obtained by Li et al. Two factors determine these differences: the first one is the scattering of the electromagnetic plane wave from the medium with a finite spatial volume, especially when the incident electromagnetic wave is obstructed by Young's pinholes. The second factor is due to the anisotropic distribution of the scattering potential of the medium, which induces the complex form of the scattered spectral density and SDOC in the far zone. Therefore, the spectral density of scattered field for the electromagnetic and anisotropic case has more complicated analytical form when comparing with Eqs. (24) and (26) of [17]. In addition, it is also found that the SDOC of scattered field derived in our paper also has different analytic form compared with Eq. (27) of [17]. This is due to the reason that the derived forms, i.e. Eqs. (39), (45) and (46) in our paper are additionally dependent of the anisotropic distributions of the strength of the scattering potential of the medium. As shown by Eq. (25), each element of the scattering potential of QH anisotropic medium can be decomposed into the product of the elements i.e. $\tilde{\mathbf{S}}_{\alpha\beta}^{(F)}$ and $\tilde{\boldsymbol{\eta}}_{\alpha\beta}^{(F)}$. So the derived results of the spectral density, SDOC and SDOP of scattered field also contain the elements of the scattering potential of QH anisotropic medium. Eqs. (39), (45) and (46), which directly result from the theoretical models introduced in [26, 27], have provided an effective approach for investigating an electromagnetic plane wave scattering upon a QH anisotropic medium. Further analysis of the spectral distribution, SDOC and SDOP of scattered field can be performed by numerical plots of Eqs. (39), (45) and (46), which do not listed in this paper to save space. The following results can be obtained for weak scattering of Young's diffractive electromagnetic waves from QH anisotropic medium:

1. For a truncated electromagnetic plane wave scattering upon a spatially QH anisotropic medium, the resultant spectral density, DOC and DOP of scattered field are dependent on the scattering path $r\hat{\mathbf{s}}$. The scattering angle $\theta$ is, thus critical for determination of the second-order statistical moments of scattered field.

2. Secondly, the spectral density, DOC and DOP of scattered field are significantly affected by the initial polarization properties of incident electromagnetic waves. The proportional factor $\gamma(\omega)$ appearing in Eqs. (39), (45) and (46) indicates the vectorial correlations of incident electromagnetic waves also influence final distributions of the spectral density, DOC and DOP of scattered field.

3. The anisotropic distributions of the spatial correlations of QH medium also have essential effects on the statistical properties of scattered field. The major factors $\boldsymbol{\Gamma}_{\alpha\alpha}(\hat{\mathbf{s}},\omega)$ and $\boldsymbol{\Gamma}_{\alpha\alpha}(\hat{\mathbf{s}}_1,\hat{\mathbf{s}}_2,\omega)$ presented in Eqs. (39), (45) and (46) incorporate both the Fourier transformations of the elements of the strength and normalized correlation coefficient of the scattering potential of QH anisotropic medium. Specifically, distributions of the spectral density, DOC and DOP of scattered field are highly dependent on $\tilde{\mathbf{S}}_{\alpha\beta}^{(F)}$ and $\tilde{\boldsymbol{\eta}}_{\alpha\beta}^{(F)}$, which have been shown in close relations to the anisotropic spatial correlations of the scattering potential, as demonstrated by Eq. (25).

4. Both the factors $\boldsymbol{\Gamma}_{\alpha\alpha}(\hat{\mathbf{s}},\omega)$ and $\boldsymbol{\Gamma}_{\alpha\alpha}(\hat{\mathbf{s}}_1,\hat{\mathbf{s}}_2,\omega)$ incorporate the ratio for spacing of pinholes to the propagation distance i.e., $d/R$, as demonstrated by Eq. (27) and (40). Essential dependence on the geometrical dimensions of truncated pinholes, then is also indicated for spectral density and DOC and DOP of scattered field. Finally variations of the spacing between Young's pinholes, or the modifications of the intermediate distance between pinholes and scatterer, separately produce fluctuations of the spectral density, DOC and DOP of scattered field.

## Conclusion

The fundamental theory of a Young's diffractive electromagnetic plane wave scattering on spatially anisotropic QH medium was introduced. Second-order moments, spectral density and DOC and DOP of scattered field were conformed to be essentially dependent on the scattering angle, the vectorial correlations of incident electromagnetic waves and the anisotropic spatial correlations of QH medium. Deduced expressions in this paper also indicate the configuration of Young's pinholes, generally, determining statistical characteristics of scattered field. Results in this paper may contribute to the future analysis of light generated from the scattering of truncated spherical waves, or solving inverse scattering problems by utilizing a laser beam obstructed by finite apertures.